\documentclass[twocolumn,showpacs,preprintnumbers,amsmath,amssymb]{revtex4}
\usepackage{bbm}
\usepackage{amsfonts}
\usepackage{color}
\usepackage{mathrsfs}
\usepackage{graphicx}
\usepackage{dcolumn}
\usepackage{bm}
\begin{document}

\preprint{APS/123-QED}
\title{Optimal Decoherence Control in non-Markovian Open, Dissipative Quantum Systems}
\author{Wei Cui}
\author{Zairong Xi}
 \email{zrxi@iss.ac.cn}
 \author{Yu Pan}%
\affiliation{%
Key Laboratory of Systems and Control, Institute of Systems Science,
Academy of Mathematics and Systems Science, Chinese Academy of
Sciences, Beijing 100080, P.~R.~China
}%

\begin{abstract}
 We investigate the optimal control problem for non-Markovian open,
dissipative quantum system. Optimal control using Pontryagin maximum
principle is specifically derived. The influences of Ohmic reservoir
with Lorentz-Drude regularization are numerically studied in a
two-level system under the following three conditions:
$\omega_0\ll\omega_c$, $\omega_0\approx\omega_c$ or
$\omega_0\gg\omega_c$, where $\omega_0$ is the characteristic
frequency of the quantum system of interest, and $\omega_c$ the
cut-off frequency of Ohmic reservoir. The optimal control process
shows its remarkable influences on the decoherence dynamics. The
temperature is a key factor in the decoherence dynamics. We analyze
the optimal decoherence control in high temperature, intermediate
temperature, and low temperature reservoirs respectively. It implies
that designing some engineered reservoirs with the controlled
coupling and state of the environment can slow down the decoherence
rate and delay the decoherence time. Moreover, we compare the
non-Markovian optimal decoherence control with the Markovian one and
find that with non-Markovian the engineered artificial reservoirs
are better than with the Markovian approximation in controlling the
open, dissipative quantum system's decoherence.
\end{abstract}


\maketitle

\section{Introduction}
The theory of open quantum systems deals with the systems that
interact with their surrounding environments
\cite{H.P.Breuer,zhang,slotine,zanzrdi,chuang,Alicki,p.zanardi}.
Such systems are of great interest, and these open quantum systems
have been extensively studied since the origin of quantum theory
\cite{J.Von. Neumann}. Despite of the noticeable progresses in the
theory, many fundamental difficulties still remain. One of the
problem is decoherence (or loss of coherence) due to the
interactions between system and environment. Recently, it received
intense considerations in quantum information and quantum
computation, where decoherence is regarded as a bottleneck to the
construction of quantum information processor
\cite{Nielsen,Mensky,zhang}. The persistence of quantum coherence is
relied on in quantum computer, quantum cryptography and quantum
teleportation. And it is also fundamental in understanding the
quantum world for the interpretation that the emergence of the
classical world from the quantum world can be seen as a decoherence
process due to the interaction between system and environment.

Various methods have been proposed to reduce this unexpected effect,
such as the quantum error-correction code \cite{shor,slotine},
error-avoiding code \cite{zanzrdi,chuang}, minimal decoherence model
\cite{Alicki}, Bang-Bang techniques \cite{p.zanardi}, quantum Zeno
effect (QZE) \cite{viola} and decoherence-free subspaces (DFS)
 \cite{elattari}. Unfortunately, all of these schemes cannot suppress the
unexpect effect successfully for accessorial conditions are needed.
Altafini \cite{altafini} pointed out that the irreversible
decohering dynamics is uncontrollable under coherent control.
Optimal control technique, which has been successfully studied in
chemical systems \cite{rabitz,levis,rabitz2} and classical systems
\cite{krotov}, has been exploited to control the quantum decoherence
\cite{Jirari, Sugny, zhang}, where an optimal control law was
designed to effectively suppress decoherence effects in Markovian
open quantum systems, dynamic coupling in the spin-boson model, and
time optimal control respectively. In this paper, we consider the
optimal decoherence control problem in non-Markovian quantum open
system.

Markovian approximation is used under the assumption that the
correlation time between the systems and environments is infinitely
short \cite{H.P.Breuer,C.W.Gardiner,zhang}. For neglecting the
memory effect, the Lindblad master equation has been built. However,
in some cases, such as quantum Brownian motion(QBM) \cite{zurek} and
a two-level atom interacting with a thermal reservoir with
Lorentzian spectral density \cite{Garraway}, an exactly analytic
description of the open quantum system dynamic is needed. Especially
in high-speed communication the characteristic time scales become
comparable with the reservoir correlation time, and in solid state
devices memory effects are typically non negligible. So it is
necessary to extensively study the non-Markovian master equation. We
briefly compare the non-Markovian dynamics (non-Markovian master
equation) with Markovian process (Markovian master equation) in
Appendix A. For details one can refer to Gardiner's book
\cite{C.W.Gardiner} or/and Breuer's \cite{H.P.Breuer}.

In this paper the focus will be on the optimal decoherence control
of non-Markovian quantum system, particularly the simplest system
possible, a two-level system governed by the time-convolutionless
(TCL) equation. We determine control fields which minimize the cost
functional suppressing the decoherence process by applying the
Pontryagin maximum principle (PMP) in Ohmic reservoir with
Lorentz-Drude regularization in the following three conditions:
$\omega_0\ll\omega_c$, $\omega_0\approx\omega_c$,
$\omega_0\gg\omega_c$, where $\omega_0$ is the characteristic
frequency of the quantum system of interest and $\omega_c$ the
cut-off frequency of Ohmic reservoir. Thus, $\omega_c\ll\omega_0$
implies that the spectrum of the reservoir does not completely
overlap with the frequency of the system oscillator and
$\omega_0\gg\omega_c$ implies the converse case. With this it is
possible to engineer different types of artificial reservoir, and
couple them to the system in a controlled way. We also compare our
results with no-control system evolution and the optimal control of
the open system with Markovian approximation. The main result of the
paper is that decoherence phenomenon can be successfully suppressed
in the $\omega_0\ll\omega_c$ case. Then this explores the coupling
of the system to engineered reservoirs \cite{Myatt, Turchette}, in
which the coupling and state of the environment are controllable.
This may pave a newly way to the realization of the first basic
elements of quantum computers.

The paper is organized as follows. We first introduce quantum
decoherence and the quantum master equation for driven open quantum
systems. In Sec. III we formulate the optimal control formalism and
deduced PMP with a minimum cost functional. Moreover, we consider
the non-Markovian two-level optimal control problem. In Sec. IV, we
numerically analyze the optimal control of decoherence to the
two-level system, and analyze the difference between Markovian
optimal control and non-Markovian optimal control from both the
system time evolution and the power spectrum. Conclusions and
prospective views are given in Sec.V.

\section{Modeling the quantum decoherence control system}
Consider a quantum system $S$ embedded in a dissipative environment
$B$ and interacting with a time-dependent classical external field,
i.e., the control field. The total Hamiltonian has the general form
\begin{equation}
\begin{array}{rcl}
H_{tot}&=&H_0+H_{B}+H_{int}\\
&=&H_S+H_{C}(t)+H_{B}+H_{int},
\end{array}
 \end{equation}
where $H_{S}$ is the Hamiltonian of the system, $H_{C}(t)$ the
Hamiltonian of the control field, and $H_{B}$ the bath and $H_{int}$
their interaction that is responsible for decoherence. The operators
$H_{S}$ and $H_{B}$ act on $\mathcal {H}_{S}$ and $\mathcal
{H}_{B}$, respectively. The operator $H_{C}(t)$ contains a
time-dependent external field to adjust the quantum evolution of the
system. One of the central goals of the theoretical treatment is
then the analysis of the dynamical behavior of the populations and
coherences, which are given by the elements of the reduced density
matrix, defined as
\begin{equation}
 \begin{split}
\rho_{S}(t)=tr_{B}[\rho_{tot}(t)],
 \end{split}
 \end{equation}
where $\rho_{tot}$ is the total density matrix for both the system
and the environment, $tr_{B}$ the partial trace taken over the
environment.  The driven model consists of a $N$-level system
interacting with a thermal bath in the presence of external control
field \cite{zhang,zhang2}, and the Hamiltonian is
 \begin{equation}
 \label{SH}
 \begin{split}
   H_C(t)=\sum_{i}u_{i}(t)H_{i}
 \end{split}
 \end{equation}
$H_{i}$ is the control Hamiltonian adjusted by the control
parameters $u_{i}(t)$, and $u_{i}(t)$ represents the control field.
The Hamiltonian of the environment is assumed to be composed of
harmonic oscillators with natural frequencies $\omega_i$ and masses
$m_i$,
 \begin{equation}
 \begin{split}
   H_{B}=\sum_{i=1}^{N}(\frac{p_i^2}{2m_i}+\frac{m_i}{2}x_i^2\omega_i^2),
 \end{split}
 \end{equation}
 where $(x_1,x_2,\cdots ,x_N,p_1,p_2,\cdots, p_N)$ are the
 coordinates and their conjugate momenta, and the Planck constant
 $\hbar$ is assigned to be $1$. The interaction Hamiltonian between the system $S$ and the environment $B$
 is assumed to be bilinear \cite{H.P.Breuer},
\begin{equation}
 \begin{split}
   H_{int}=\alpha\sum_{n}A_{n}\otimes B_{n}.
 \end{split}
 \end{equation}

The interaction Hamiltonian in the interaction picture therefore
takes the form
\begin{equation}
\begin{array}{rcl}
H_{int}(t)&=&e^{i(H_S+H_B)t}H_{int}e^{-i(H_S+H_B)t}\\
&=&\alpha\sum_nA_n(t)\otimes B_n(t),
\end{array}
 \end{equation}
where \[
\begin{array}{rcl}
A_n(t)&=&e^{iH_St}A_ne^{-iH_St},\\
B_n(t)&=&e^{iH_Bt}B_ne^{-iH_Bt}.
\end{array}\]

The
 effect of the environment on the dynamics of the system can be
seen
 as a interplay between the dissipation and fluctuation phenomena. And it
 is the general environment that makes the quantum system loss of
 coherence (decoherence). In general the decoherence can be
 demonstrated as the interaction between the system and environment. Then the
 reduced density matrix of the system can evolve into the form
\begin{equation}
 \begin{split}
\rho_{T}\simeq\sum_{n}|c_{n}|^2|a_{n}\rangle\langle a_{n}|
 \end{split}
 \end{equation}
which describes a statistical mixture of noninterfering states.
Thus, a commonly proposed way to analyze decoherence is by examining
how the nondiagonal elements of the reduced density matrix evolve
under the master equation.

 In the present work, we shall concentrate on
optimal control of the decoherence effect in open quantum system.
The kinetic equation of strong coupling non-Markovian quantum system
is the following exact time-convolutionless (TCL) form of the master
equation,
\begin{equation}
\begin{split}
\frac{d}{dt}\mathcal {P}\rho(t)=\sum_iu_i\tilde{\mathcal
{K}}_i(t)\mathcal {P}\rho(t)+\mathcal {K}(t)\mathcal
{P}\rho(t)+\mathcal {I}(t)\mathcal {Q}\rho(t_0),
   \end{split}
\end{equation}
with the time-local generator, called the TCL generator
\begin{equation}
\begin{split}
\tilde{\mathcal {K}}_i(t)=e^{iH_St}H_ie^{-iH_St},~~~~
 \mathcal {K}(t)=\alpha\mathcal
{P}\mathcal {L}(t)[1-\Sigma(t)]^{-1}\mathcal {P},
   \end{split}
\end{equation}
and the inhomogeneity
\begin{equation}
\begin{split}
\mathcal {I}(t)=\alpha\mathcal {P}\mathcal
{L}(t)[1-\Sigma(t)]^{-1}g(t,t_0)\mathcal {Q},
   \end{split}
\end{equation}
where $\Sigma(t)$ is the superoperator
$$\Sigma(t)=\alpha\int_{t_0}^tds \mathcal {G}(t,s)\mathcal {Q}\mathcal {L}(s)\mathcal {P}G(t,s).$$
 For detail one see Appendix A
and/or \cite{H.P.Breuer}

  In order to facilitate the calculations, we will convert the
  differential equation (8) from the complex density matrix
  representation into the so-called coherent vector representation
  \cite{Blum, altafini, zhang}. Firstly, we choose an orthonormal basis of
  $N\times N$ matrices $\{(I, \Omega_{j})\}_{j=1,2,\cdots, N^2-1}$ with
  respect to the inner product $\langle X, Y\rangle=tr(X^{\dag}Y)$,
  where $I$ is the N-dimensional identity matrix and $\Omega_j$ are
  $N\times N$ Hermitian traceless matrices. In particular, the
  Hermitian density matrix $\rho$ can be represented as
  $\rho=\frac{1}{N}I+\sum_{i}x_i\cdot\Omega_i$, where $\vec{x}=(x_1, x_2, \cdots
  x_{N^2-1})^{T}$ is a real $(N^2-1)$ dimensional vector, called the coherent vector of $\rho$. This is the well-known Bloch
  vector representation of quantum systems. Thus the master
  equation (8) can be rewritten as a differential equation of the
  coherent vector:
\begin{equation}
\begin{array}{rcl}
\label{Syst}
\dot{x}(t)&=&O_0x(t)+\sum_{i=1}^{k}u_{i}(t)O_{i}x(t)+L_1(t)x(t)+L_2(t),\\
 \end{array}
 \end{equation}
 with the initial condition,
 $$x(t_0)=x_0$$
 where $O_0, O_i\in \text{so}(N^2-1)$ are the adjoint representation matrices of $-iH_0, -iH_i$ respectively,
 and $x_0$ is the coherence vector of $\rho_0$, and the term
 $L_1(t)x(t)$ represents the decoherence process, $k$ is the
 number of control fields, $\sum_i^ku_iH_i$ adjusts the quantum evolution such that the coherence
 is conserved.

\section{quantum optimal control problem}
\subsection{General Formalism}
As well known, the evolution of the state variable $x(t)$ governed
by the master equation (\ref{Syst}) depends not only on the initial
state $x_0$ but also on the choice of the time-dependent control
variable $u(t)$. Some earlier works to these control problem are
listed in the reference \cite{Thorwart,Grifoni,Shao}. Especially,
the exact result was considered of the quantum two-state dynamics
driven by stationary non-Markovian discrete noise in \cite{Goychuk}.
In this section, we are going to suppress the unexpected effect of
decoherence by optimal control technique that wants to force the
system evolving along some prescribed cohering trajectories. The
target state chosen is the free evolution of the closed system:
\begin{equation}
 \begin{split}
\dot{\rho}_T(t)=-i[H_0, \rho_T(t)],
 \end{split}
 \end{equation}
which is equivalent to $x^0(t)=e^{O_0(t-t_0)}x_0$. The cost
functional is
\begin{equation}
 \begin{split}
J[u(t)]=\Psi[x(t_f),x^0(t_f)]+\int_{t_o}^{t_f}\Theta(x(t),x^0(t_f),u(t))dt,
 \end{split}
 \end{equation}
where the functional $\Psi[x(t_f),x^0(t_f)]$ represents distance
between the system and objects at final time and the functional
$\int_{t_o}^{t_f}\Theta(x(t),x^0(t_f),u(t))$ accounts for the
transient response with $\Theta(x(t),x^0(t_f),u(t))\geq 0$.

 The
optimal control problem considered in this paper is to minimize the
cost functional $J[u(t)]$ with some dynamical constraints. That is,
our problem is
\begin{equation}
\begin{array}{rcl}
\min_{u\in\mathcal{U}_{[t_0,t_f]}} J[u(t)]
&=&\Psi[x(t_f),x^0(t_f)]+\int_{t_o}^{t_f}\Theta(t)dt,\\
\dot{x}(t)&=&O_0x(t)+\sum_{i=1}^{k}u_{i}(t)O_{i}x(t)\\&+&L_1(t)x(t)+L_2(t),\\
x(t_0)&=&x_0, ~t\in[t_0, t_f],
\end{array}
\end{equation}
where $\mathscr{U}_{[t_0,t_f]}=\{u(\cdot):[t_0,t_f]\longrightarrow
\mathbb{R}^k\}$ and $u(\cdot)$ piecewise continuous.

 Using the
Pontryagin's maximum principle \cite{krotov}, the optimal solution
to this problem is characterized by the following
Hamilton-Jacobi-Bellman(HJB) Equation
\begin{widetext}
 \begin{equation}
 \begin{cases}
 \frac{\partial J}{\partial t}+min_{u\in\mathscr{U}_{[t_0,t_f]}}\{O_0x(t)+\sum_{i=1}
 ^{k}u_{i}(t)O_{i}x(t)+L_1(t)x(t)+L_2(t)+\Theta(x(t),x^0(t_f),u(t))\}=0,\\
 J(x(t_f),t_f)=\Psi[x(t_f)].
 \end{cases}
 \end{equation}
 \end{widetext}
In general, it is usually difficult to obtain the analytic solution.
Nevertheless, one can always have numerical solution. To illustrate
this method and give more insight, we will consider this problem for
the non-Markovian two-level system in the following.
\subsection{Optimal Control of non-Markovian Two-Level System}
In this subsection we consider the decoherence of two-level system
whose controlled Hamiltonian is
\begin{equation}\label{SHP}H_0=\frac{1}{2}\{\omega_0\sigma_z+u_x(t)\sigma_x+u_y(t)\sigma_y\},
\end{equation}
where $\sigma_k$ with $k=x,y,z$ are the Pauli matrices;
 $\omega_0$ is the
transition frequency of the two-level system, and $u(t)$ is the
modulation by the time-dependent external control field. In fact,
the free Hamiltonian is $H_S=\frac{1}{2}\omega\sigma_z$. Then the
control Hamiltonian can be described by $\sigma_x, \sigma_y$
according to Cartan decomposition of the Lie algebra $\text{su}(2)$,
which was discussed by Zhang et. al in details \cite{zhang2}. This
is the standard model for atom-field interaction \cite{scully,
meystre, Anastopoulos, Shresta}.

In our two-level system the assumed bilinear interaction between the
system $S$ and the environment $B$ can be written as
\begin{equation}
 H_{int}=\alpha\left(\sigma_{+}\otimes B+\sigma_{-}\otimes B^{\dag}\right)~~~\text{with}~~ B=\sum_ik_ia_i,
 \end{equation}
where $\sigma_{\pm}=(\sigma_x\pm i\sigma_y)/2$, the raising and
lowering operator respectively, $k_i$ is the coupling constant
between the spin coordinate and the $i$th environment oscillator,
and $a_i$ is the annihilation operator of the $i$th harmonic
oscillators of the environment. The coupling constants enter the
spectral density function $J(\omega)$ of the environment defined by
\begin{equation}
J(\omega)=\frac{\pi}{2}\sum_i\frac{k_i}{m_i\omega_i}\delta(\omega-\omega_i)
 \end{equation}
 and the index $i$ labels the different field models of
the reservoir with frequencies $\omega_i$. In the
 continuum limit the spectral density has the form
\begin{equation}
 \begin{split}
J(\omega)=\eta\omega(\frac{\omega}{\omega_c})^{n-1}\exp(-\frac{\omega}{\omega_c}),
 \end{split}
 \end{equation}
 where $\omega_c$ is a cutoff frequency, and $\eta$ a dimensionless
 coupling constant. The environment is classified as Ohmic, sub-Ohmic, and sup-Ohmic according to $n=1$,
$0<n<1$, and $n>1$, respectively \cite{An,
 Leggett, weiss}.

In this case, the open quantum system can be written as follows
\cite{H.P.Breuer,Maniscalco,Maniscalco2}
\begin{equation}
\label{DOS}
\begin{array}{rcl}
\dot{\rho}_S=-\frac{i}{2}\omega_0[\sigma_z,
\rho_S]-\frac{i}{2}u_x(t)[\sigma_x,
\rho_S]-\frac{i}{2}u_y(t)[\sigma_y, \rho_S]\\
+\frac{\Delta(t)+\gamma(t)}{2}\{2\sigma_{-}\rho_S\sigma_{+}-\sigma_{+}\sigma_{-}\rho_S-\rho_S\sigma_{+}\sigma_{-}\}\\+
\frac{\Delta(t)-\gamma(t)}{2}\{2\sigma_{+}\rho_S\sigma_{-}-\sigma_{-}\sigma_{+}\rho_S-\rho_S\sigma_{-}\sigma_{+}\}.
\end{array}
\end{equation}

 For convenience we map the density
matrix of the two-level system onto the Bloch vector
$x(t)=(x_1(t),x_2(t),x_3(t))^{T}\in \mathbb{R}^3$ defined by
$x(t)=Tr[\sigma\rho(t)]$, which implies that
\begin{equation}
\begin{array}{rcl}
x_1(t)&\equiv&\rho_{01}(t)+\rho_{10}(t),\\
x_2(t)&\equiv&i(\rho_{01}(t)-\rho_{10}(t)),\\
x_3(t)&\equiv&\rho_{00}(t)-\rho_{11}(t) .
\end{array}
\end{equation}

Then the explicit equations of motion for the components of the
Bloch vector read
\begin{equation}
\label{TLS}\left\{
 \begin{array}{rcl}
\dot{x_1}(t)&=&-\Delta(t)x_1(t)-\omega_0x_2(t)+x_3(t)u_y(t),\\
\dot{x_2}(t)&=&\omega_0x_1(t)- \Delta(t)x_2(t)-x_3(t)u_x(t),\\
\dot{x_3}(t)&=&-2\Delta(t)x_3(t)-2\gamma(t)+x_2(t)u_x(t)-x_1(t)u_y(t),
\end{array}\right.
 \end{equation}
where the expressions for the relevant time dependent coefficients,
up to the second order in the system-reservoir coupling constant,
are given by \cite{Maniscalco, H.P.Breuer}
\begin{equation}
\label{DDC}
 \begin{array}{rcl}
\Delta(t)&=&\int_0^td\tau k(\tau)\cos(\omega_0\tau)\\
\gamma(t)&=&\int_0^td\tau \mu(\tau)\sin(\omega_0\tau)
 \end{array}
 \end{equation}
with
\begin{equation}
 \begin{array}{rcl}
k(\tau)&=&2\int_0^{\infty}d\omega
J(\omega)\coth[\hbar\omega/2k_BT]\cos(\omega \tau),\\
\mu(\tau)&=&2\int_0^{\infty}d\omega J(\omega)\sin(\omega \tau),
 \end{array}
 \end{equation}
being the noise and the dissipation kernels, respectively. The
equation (\ref{TLS}) can be written compactly as

\begin{equation}
 \begin{split}
\dot{x}(t)=A(t)x(t)+B(t),
 \end{split}
 \end{equation}
where
\[
A(t)=\left(\begin{array}{ccc}
-\Delta(t)&-\omega_0&u_y(t)\\
\omega_0&-\Delta(t)&-u_x(t)\\
-u_y(t)&u_x(t)&-2\Delta(t)
\end{array}\right)
\]
and
\[
B(t)=\left(\begin{array}{ccc}
0\\
0\\
-2\gamma(t)
\end{array}\right).
\]

Let the Ohmic spectral density with a
 Lorentz-Drude cutoff function,
\begin{equation}
 \begin{split}
J(\omega)=\frac{2\gamma_0}{\pi}\omega\frac{\omega_c^2}{\omega_c^2+\omega^2},
 \end{split}
 \end{equation}
 where $\gamma_0$ is the frequency-independent damping constant and usually assumed to be $1$.
 $\omega$ is the frequency of the bath, and $\omega_c$ is the high-frequency cutoff.
For this type of spectral density the bath correlations can be
determined analytically as
\begin{equation}
 \begin{split}
k(\tau)=4
k_BT{\omega_c}^2\sum_{n=-\infty}^{+\infty}\frac{\omega_ce^{-{\omega_c}|\tau|}-|\nu_n|e^{-|\nu_n||\tau|}}{\omega_c^2-\nu_n^2}
 \end{split}
 \end{equation}
where $\nu_n=2\pi n k_BT$ and
\begin{equation}
 \begin{split}
\mu(\tau)=2\hbar\omega_c^2e^{-\omega_c|\tau|} sign~ \tau.
 \end{split}
 \end{equation}
 Then the analytic expression for the dissipation coefficient $\gamma(t)$ appearing in the equation
(\ref{DDC}) is
 \begin{equation}
 \begin{split}
 \gamma(t)=\frac{\alpha^2\omega_0r^2}{1+r^2}[1-e^{-r\omega_0t}\cos(\omega_0t)-re^{-r\omega_0t}\sin(\omega_0t)],
 \end{split}
 \end{equation}
and the closed analytic expression for $\Delta(t)$ is
\cite{Maniscalco2}
\begin{widetext}
\begin{equation}
\label{Delta}
\begin{array}{rcl}
\Delta(t)&=&\alpha^2\omega_0\frac{r^2}{1+r^2}\{\coth(\pi
r_0)-\cot(\pi
r_c)e^{-\omega_ct}[r\cos(\omega_0t)-\sin(\omega_0t)]+\frac{1}{\pi
r_0}\cos(\omega_0t)[\bar{F}(-r_c,t)\\&&+\bar{F}(r_c,t)-\bar{F}(ir_0,t)-\bar{F}(-ir_0,t)]-\frac{1}{\pi}
\sin(\omega_0t)[\frac{e^{-\nu_1t}}{2r_0(1+r_0^2)}[(r_0-i)\bar{G}(-r_0,t)\\&&+(r_0+i)\bar{G}(r_0,t)]+\frac{1}{2r_c}[\bar{F}(-r_c,t)-\bar{F}(r_c,t)]]\},
\end{array}
\end{equation}
\end{widetext}
where $r_0=\omega_0/2\pi k_BT$, $r_c=\omega_c/2\pi k_BT$,
$r=\omega_c/\omega_0$, and
 \begin{equation}
 \begin{split}
\bar{F}(x,t)\equiv _2F_1(x,1,1+x,e^{-\nu_1t}),
 \end{split}
 \end{equation}
  \begin{equation}
 \begin{split}
\bar{G}(x,t)\equiv _2F_1(2,1+x,2+x,e^{-\nu_1t}).
 \end{split}
 \end{equation}
 $ _2F_1(a,b,c,z)$ is the hypergeometric function and takes the form
\begin{eqnarray*}
_2F_1(a,b,c,z)&=&1+\frac{ab}{1!c}z+\frac{a(a+1)b(b+1)}{2!c(c+1)z^2}z^2+\cdots\\
&=&\sum_{n=0}^{\infty}\frac{(a)_n(b)_n}{(c)_n}\frac{z^n}{n!},
\end{eqnarray*}
where $(a)_n$ is a Pochhammer symbol. Under the high temperature
limit, we have
 \begin{equation}
 \begin{split}
\Delta(t)=2\alpha^2\kappa
T\frac{r^2}{1+r^2}\{1-e^{-r\omega_0t}[\cos(\omega_0t)-\frac{1}{r}\sin(\omega_0t)]\}.
 \end{split}
 \end{equation}

In the following we consider the optimal control formalism of our
two-level system. For simplicity we define the cost functional as:
 \begin{equation}
 \begin{split}
J[u(t)]=\int_{t_{0}}^{t_{f}}[(x(t)-x^0(t))^2+\theta u^{T}(t)u(t)]dt
 \end{split}
 \end{equation}
where $\theta > 0$ is a weighting factor used to achieve a balance
between the tracking precision and the control constraints. The
corresponding Hamiltonian function is
\begin{widetext}
\begin{eqnarray*}
\mathcal{H}(x(t),u(t),\lambda(t),t)&=&[(x(t)-x^0(t))^2+\theta
u^{T}(t)u(t)]+\lambda(t)^{T}[A(t)x(t)+B(t)]\\
&=&[(x_1(t)-x_1^0(t))^2+(x_2(t)-x_2^0(t))^2+(x_3(t)-x_3^0(t))^2+\theta(u_1^2(t)+u_2^2(t))]\\
&&+\lambda_1(t)[-\Delta(t)x_1(t)-\omega_0x_2(t)+x_3(t)u_y(t)]+\lambda_2(t)[\omega_0x_1(t)-\Delta(t)x_2(t)\\
&&-x_3(t)u_x(t)]+\lambda_3(t)[-2\Delta(t)x_3(t)-2\gamma(t)+x_2(t)u_x(t)-x_1(t)u_y(t)],
\end{eqnarray*}
\end{widetext}
 where $\lambda(t)=(\lambda_1(t),\lambda_2(t),\lambda_3(t))^{T}$ is the so-called Lagrange
 multiplier and $x^0(t)=(x^0_1(t),x^0_2(t),x^0_2(t))$ is the target trajectory defined by $\dot{\rho}=-\frac{i}{2}[H_0,
 \rho]$. It is easy to see that $x^0(t)=(x^0_1\cos\omega t-x^0_2\sin\omega t, x^0_1\sin\omega t+x^0_2\cos\omega,
 x^0_3)$.
 The optimal solution can be solved by the following
 differential equation with two-sided boundary values,
 \begin{equation}
 \begin{cases}
 \dot{x}^{*}(t)=\frac{\partial \mathcal{H}}{\partial
 \lambda}=A(t)x(t)+B(t)
 \\\dot{\lambda}(t)=-\frac{\partial \mathcal{H}}{\partial
 x}=-2[x(t)-x^0(t)]-A(t)^T\lambda(t)
 \\x^*(0)=x_0
\\\lambda(t_f)=0
 \end{cases}
 \end{equation}
 together with
 \begin{equation}
 \begin{cases}
 \frac{\partial \mathcal{H}}{\partial u}|_*=\frac{\partial \mathcal{H}(x^*(t),u^*(t),\lambda(t),t)}{\partial
 u}=0,
 \\\frac{\partial^2\mathcal{H}}{\partial u^2}|_*=\frac{\partial^2\mathcal{H}(x^*(t),u^*(t),\lambda(t),t)}{\partial
 u^2}\leq0,
  \end{cases}
 \end{equation} which implies that
 \begin{equation}
 \begin{cases}
 u_x(t)=\frac{1}{2\theta}\{\lambda_2x_3-\lambda_3x_2\},\\
 u_y(t)=\frac{1}{2\theta}\{\lambda_3x_1-\lambda_1x_3\}.
  \end{cases}
 \end{equation}

 The minimum principle requires the solution of the complicated
 nonlinear equations. When there is one and only one solution
 $\{x(t),~\lambda(t)\}$ it is the required optimal solution \cite{krotov}.
 In general, it is difficult to obtain the analytic solution, if possible existence,
 to the above optimal control
 problem. So numerical demonstration to this problem will be considered
 in the next section.

\section{numerical demonstration and discussions}

In this section, we use the formalism of the preceding section to
determine the optimal control of the decoherence. Though the
spin-bath models of real systems are expected to be more complicated
than the two-level Hamiltonians considered here, we study the system
in various aspects to understand the effect of this simple system on
the decoherence control.

In our simulations, the system parameters are chosen as following,
$x(0)=(\frac{\sqrt{3}}{2},\frac{-\sqrt{2}}{4},\frac{-\sqrt{2}}{4})$,
strong coupling constant $\alpha^2=0.01$, weighting factor
$\theta=1$, $\omega_0=1$ as the norm unit. Moreover, we regard the
temperature as a key factor in decoherence process. For high
temperature $k_BT=300\omega_0$, intermediate temperature
$k_BT=3\omega_0$, and low temperature $k_BT=0.3\omega_0$. Another
reservoir parameter playing a key role in the dynamics of the system
is the ratio $r=\omega_c/\omega_0$ between the reservoir cutoff
frequency $\omega_c$ and the system oscillator frequency $\omega_0$.
As we will see in this section, by varying these two parameters
$k_BT$ and $r=\omega_c/\omega_0$, both the time evolution and the
optimal control of the open system vary prominently from Markovian
to non-Markovian.

\subsection{High temperature reservoir}
For high reservoir temperature, diffusion coefficient $\Delta(t)$
(30) has the approximation form (33), which plays a dominant role
since $\Delta(t)\gg\gamma(t)$. Note that, for time $t$ large enough,
the coefficients $\Delta(t)$ and $\gamma(t)$ can be approximated by
their Markovian stationary values
$\Delta_M=\Delta(t\rightarrow\infty)$ and
$\gamma_M=\gamma(t\rightarrow\infty)$. From eqs.(29) and (30) we
have
\begin{equation}
\gamma_M=\frac{\alpha^2\omega_0r^2}{1+r^2},
 \end{equation}
 and \begin{equation}
\Delta_M=\alpha^2\omega_0\frac{r^2}{1+r^2}\coth(\pi r_0).
 \end{equation}
Then, under high temperature, noting
 \[\coth(\pi r_0)\simeq1+\frac{1}{\pi
 r_0}\simeq\frac{2kT}{\omega_0},\]
  \begin{equation}
\Delta_M^{HT}=2\alpha^2kT\frac{r^2}{1+r^2}.
 \end{equation}
Inserting Eqs.(38) and (40) into Eqs.(35) one can easily get the
Markovian optimal decoherence control.

\begin{figure*}
\centerline{\scalebox{0.8}[0.4]{\includegraphics{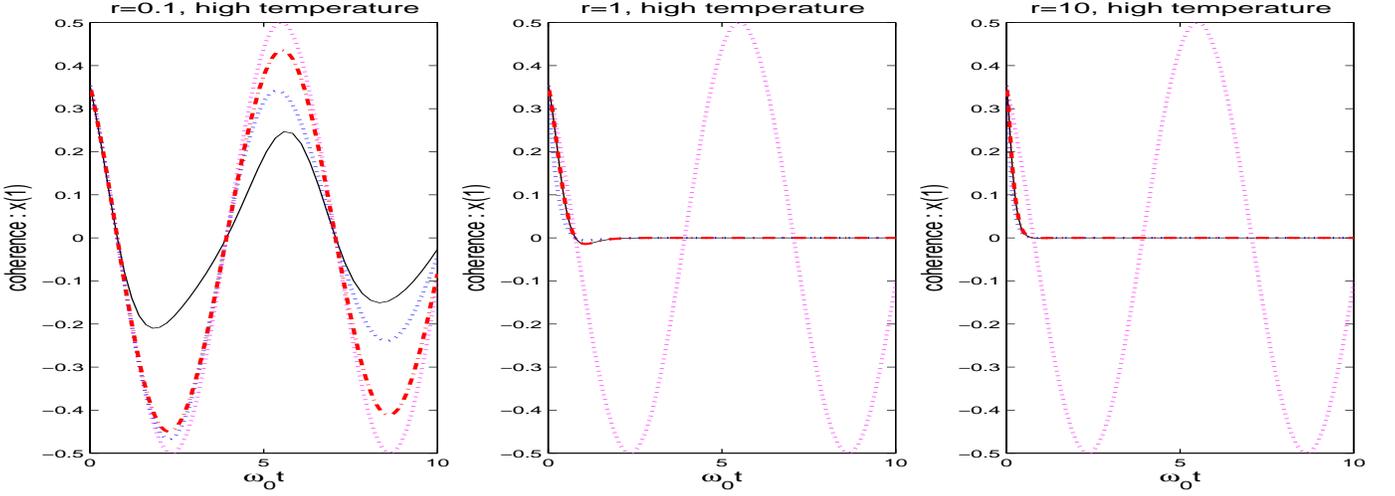}}}
\caption{(Color online)Surviving coherence in off-diagonal matrix
elements vs time $t$ [Eq.35] under high temperature environment,
without control action(black solid line), Markovian optimal
control(blue dashed line), non-Markovian optimal control(red dotted
line), and target trajectory(crimson dash-dotted line) at $r=0.1$,
$r=1$, $r=10$ respectively.}
\end{figure*}

Figure 1 shows optimal control of decoherence for
$r\ll1,~~r=1$,~~and $r\gg1$ in high temperature reservoir. All of
these contain solid line for free evolution, dashed line for
Markovian optimal control, dotted line for non-Markovian optimal
control, and dash-dotted line for target trajectory. We can see
clearly that the decoherence can be controlled perfectly in $r\ll1$
reservoir. From Figure 2 we can see that the decoherence time
$\tau_{D}$ can be delayed for a long time and its amplitude
amplified heavily with the non-Markovian control. On the other hand,
Figure 3 shows that the non-Markovian control field is changed more
rapidly than the Markovian control field and the frequency of
non-Markovian is more plenty than the Markovian, which helps to
understand that the non-Markovian case is done better than the
Markovian case and implies that it is necessary to consider the
non-Markovian case.

\begin{figure*}
\centerline{\scalebox{0.4}[0.4]{\includegraphics{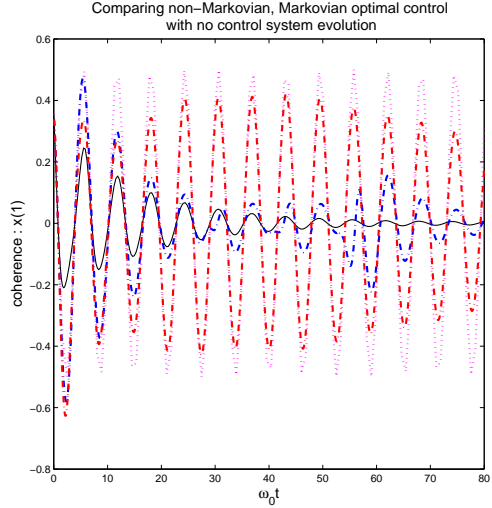}}}
\caption{(Color online)Comparing Markovian optimal control,
non-Markovian optimal control with no control under high temperature
environment, without control action(black solid line), Markovian
optimal control(blue dashed line), non-Markovian optimal control(red
dotted line), and target trajectory(crimson dash-dotted line) at
$r=0.1$.}
\end{figure*}

\begin{figure*}
\centerline{\scalebox{0.8}[0.3]{\includegraphics{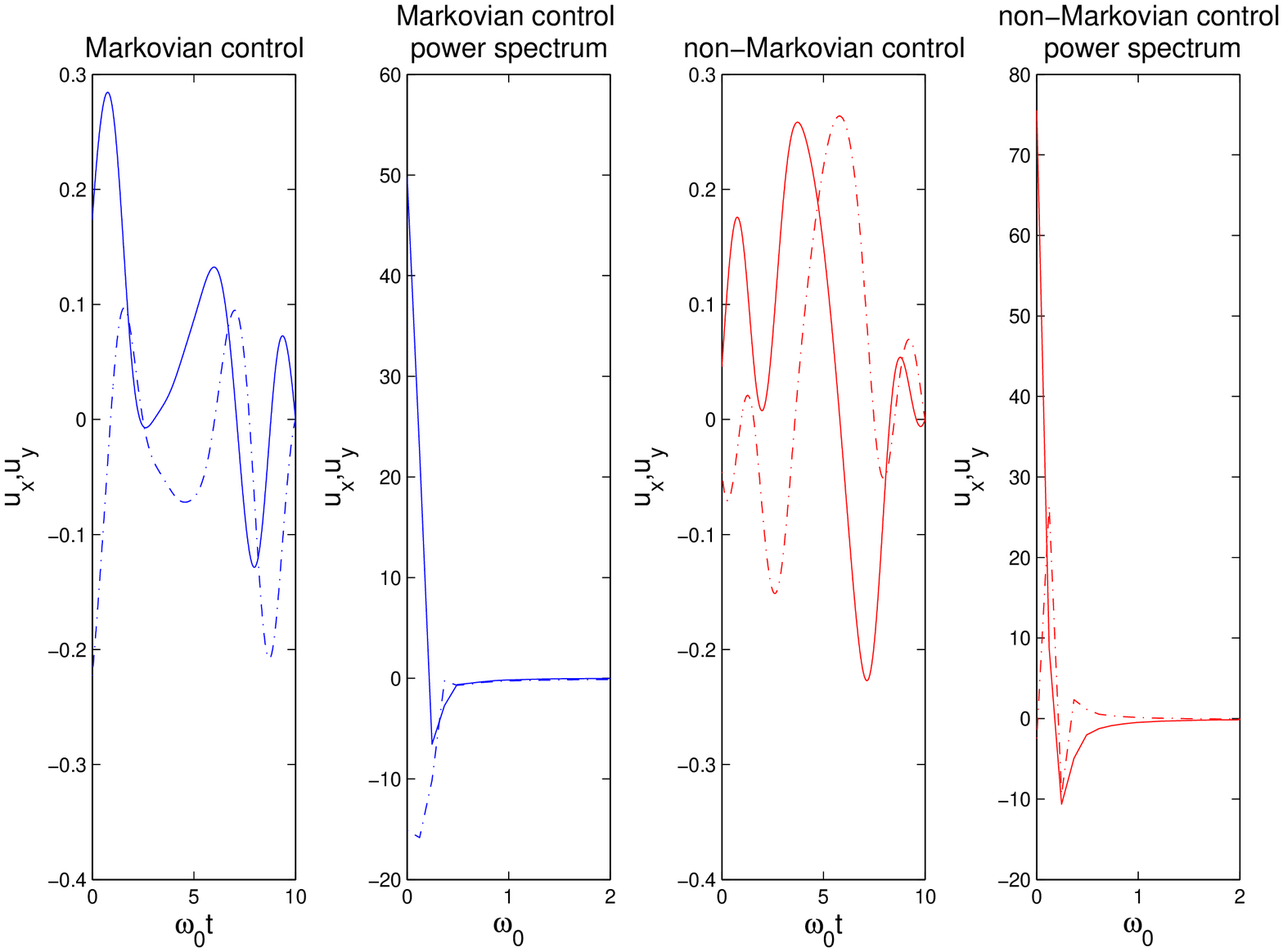}}}
\caption{(Color online)Comparing Markovian optimal control with
non-Markovian one in r=0.1 high temperture reservoir. Markovian
optimal control $u_x$(blue solid line) $u_y$(blue dash-dotted line)
, non-Markovian optimal control $u_x$(red solid line) $u_y$ (red
dash-dotted line).}
\end{figure*}

From Figure 1 we can also see that either Markovian or non-Markovian
optimal control cannot do well when $r=1$ or $r=10$. As we discussed
before, diffusion is always dominant under the high temperature. In
the case $r\ll1$, $\Delta(t)>0$ is always true \cite{Maniscalco2}.
However, Maniscalco, et. al. \cite{Maniscalco2} showed that if
$r>0.27$ the diffusion coefficient $\Delta(t)<0$, and the system
becomes non-Lindblad. It implies that the environment induced
fluctuations will be large enough. So our control field is
negligible when comparing with the high-frequency harmonic
oscillators of the reservoir.

\subsection{Lower temperature reservoir}
As decreasing temperature, the amplitude of $\Delta(t)$ becomes
smaller and smaller and $\gamma(t)$ becomes larger and larger, which
is not negligible anymore. There exists a time which relate to both
the temperature and the ratio such that after the time the
combination of dissipation and diffusion coefficient
$\Delta(t)-\gamma(t)<0$, which changes the properties of the control
system (35).

Figure 4 shows the non-Markovian optimal control and Figure 5 their
power spectrum for intermediate temperature, and Figure 6 and Figure
7 for low temperature. At intermediate temperature the non-Markovian
optimal control plays little role especially in Figure 4(b)and
Figure 4(c). Note that in Figure 4(a) and 6(a) the free evolution is
with little decoherence. We note that the optimal control does well
at low temperature in Figure 6. In Figure 6, both Markovian and
non-Markovian play an important role in controlling the decoherence
in both $r=1$ and $r=10$. They can make the quantum coherence
persistence for a long time.

\begin{figure*}
\centerline{\scalebox{0.8}[0.4]{\includegraphics{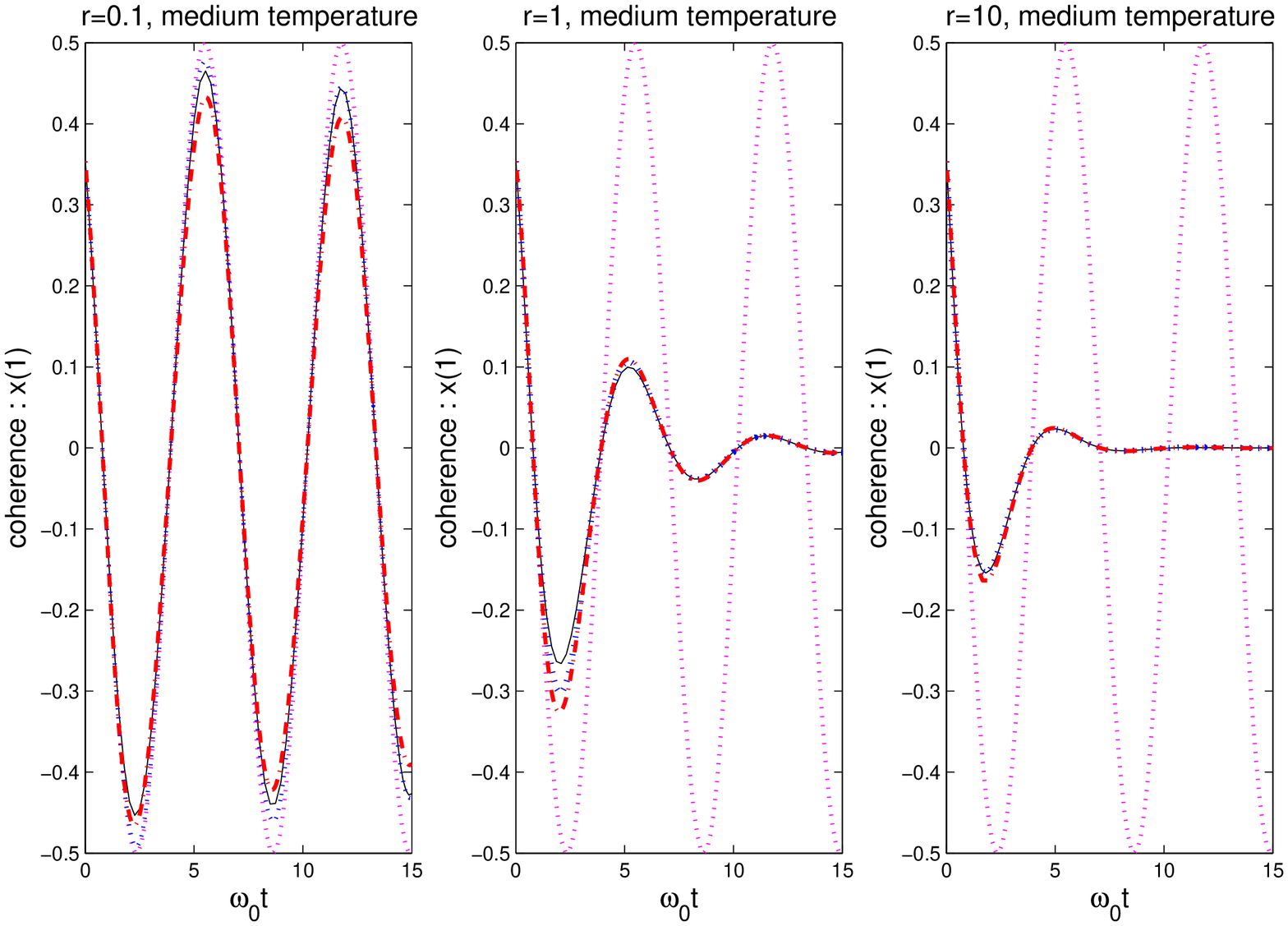}}}
\caption{(Color online)Surviving coherence in off-diagonal matrix
elements vs time $t$ [Eq.35] under medium temperature environment,
without control action(black solid line), Markovian optimal
control(blue dashed line), non-Markovian optimal control(red dotted
line), and target trajectory(crimson dash-dotted line) at $r=0.1$,
$r=1$, $r=10$ respectively.}
\end{figure*}

\begin{figure*}
\centerline{\scalebox{0.70}[0.55]{\includegraphics{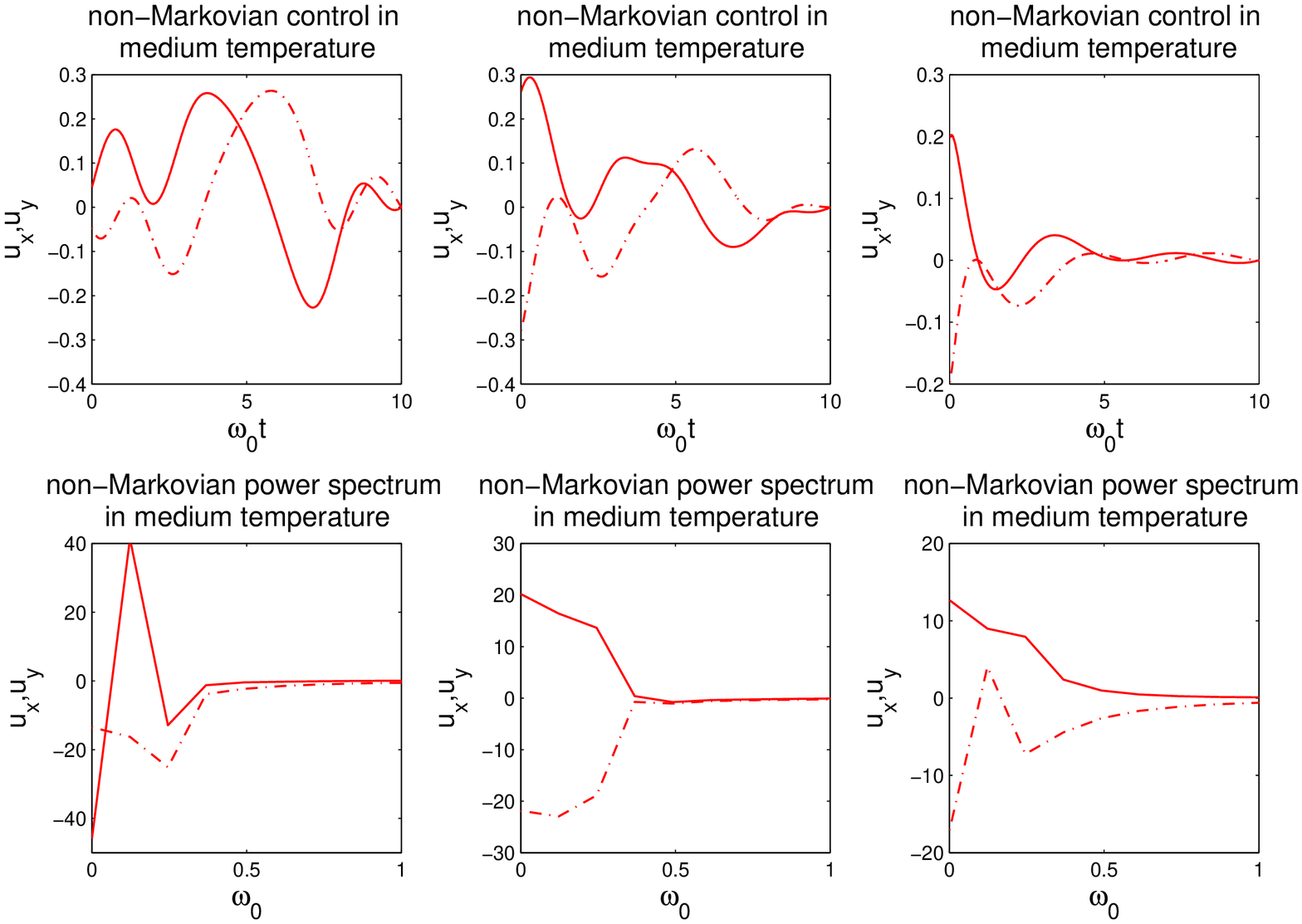}}}
\caption{(Color online)non-Markovian optimal controls and their
power spectrum in medium temperature reservoir for $r=0.1$, $r=1$,
and $r=10$ respectively. Non-Markovian optimal control $u_x$(red
solid line) $u_y$(red dash-dotted line).}
\end{figure*}

\begin{figure*}
\centerline{\scalebox{0.7}[0.55]{\includegraphics{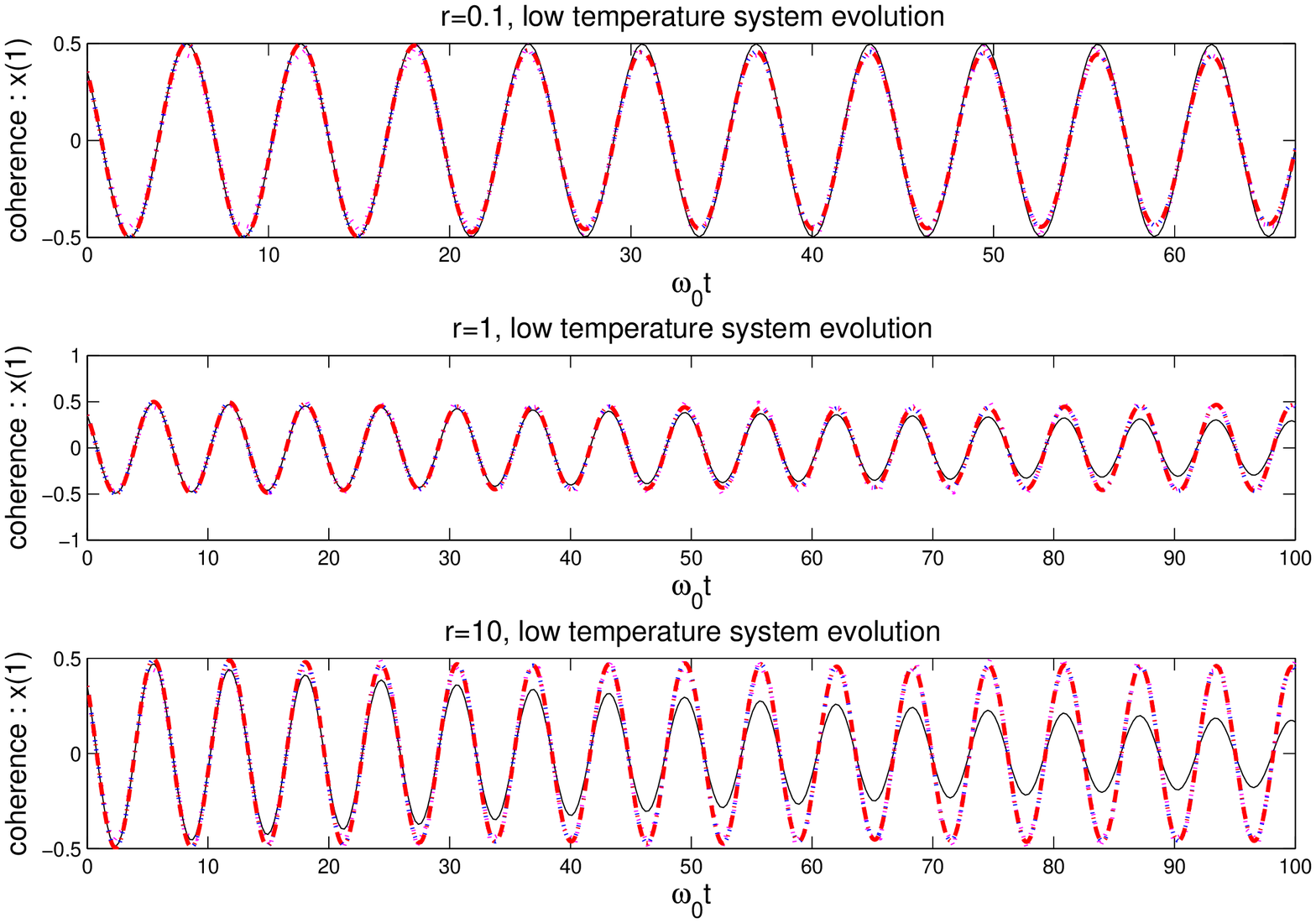}}}
\caption{(Color online)Surviving coherence in off-diagonal matrix
elements vs time $t$ [Eq.35] under low temperature environment,
without control action(black solid line), Markovian optimal
control(blue dashed line), non-Markovian optimal control(red dotted
line), and target trajectory(crimson dash-dotted line) at $r=0.1$,
$r=1$, $r=10$ respectively.}
\end{figure*}

\begin{figure*}
\centerline{\scalebox{0.7}[0.55]{\includegraphics{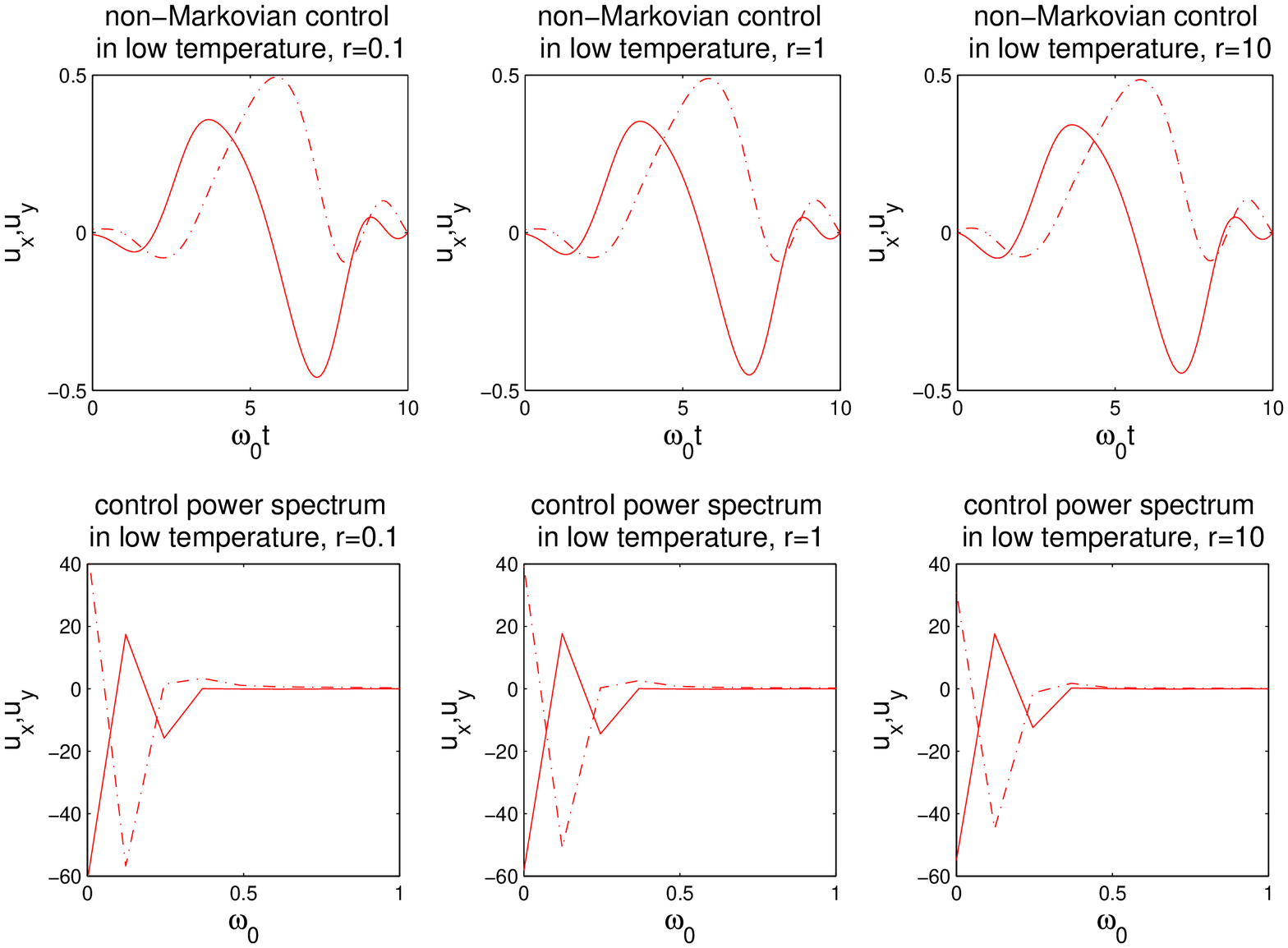}}}
\caption{(Color online)non-Markovian optimal controls and their
power spectrum in low temperature reservoir for $r=0.1$, $r=1$, and
$r=10$ respectively. Non-Markovian optimal control $u_x$(red solid
line) $u_y$(red dash-dotted line).}
\end{figure*}

\subsection{Engineering Reservoirs}
During the last two decades, great advances in laser cooling and
trapping experimental techniques have made it possible to trap a
single ion and cool it down to very low temperature. These cold
trapped ions are the favorite candidates for a physical
implementation of quantum computers and realization of the quantum
cryptography and quantum teleportation. All of these rely on the
persistence of quantum coherence. \cite{Myatt, Turchette} are the
recent experimental procedures for engineering artificial
reservoirs. They showed how to couple a properly engineered
reservoirs with the trapped atomic ion's harmonic motion. They
measured the decoherence of superpositions of coherent states and
two-Fock-state superpositions in the engineering artificial
reservoirs. Several type of engineering artificial reservoirs are
simulated, e.g., a high-temperature amplitude reservoir, a
zero-temperature amplitude reservoir, and a high-temperature phase
reservoir.

From above discussions we find that our optimal decoherence control
fields do well in the engineering artificial reservoirs.

\centerline{
 TABLE I. Controllability}
\begin{center}
\begin{tabular}{|c|c|c|c|}
\hline \multicolumn{1}{|c|}{T\&r}
 &\multicolumn{1}{c|}{Low T}
&\multicolumn{1}{c|}{Med T} &\multicolumn{1}{c|}{High T}\\ \hline
r=0.1&slow decay &slow decay &controllable(non)\\
\hline

r=1&controllable& uncontrollable   &  uncontrollable   \\
\hline

r=10&controllable& uncontrollable & uncontrollable
\\ \hline
\end{tabular}
\end{center}

Table I shows the controllable property of non-Markovian open,
dissipative quantum system.
 When $r\ll1$ the
system free evolution is with little decoherence at low and
intermediate temperature and our optimal control plays an important
role in controlling the decoherence phenomenon at high temperature.
Moreover, when $r\gg1$ and $r=1$ our optimal control also plays an
important role in controlling the decoherence phenomenon at low
temperature. They indicate that these engineered reservoirs could be
designed that the coupling and state of the environment can be
controlled to slow down the decoherence rate and delay decoherence
time.

\section{conclusions}
In the present work, we have studied the optimal control of the
decoherence for the non-Markovian open quantum system. In the
general formalism we proposed the optimal control problem and
derived the corresponding Hamilton-Jacobi-Bellman equation. Usually
this kind of problem is difficult to be analytically solved. Then we
considered this problem in the non-Markovian two-level system.
Through transforming its master equation into the Bloch vector
representation we obtained the corresponding differential equation
with two-sided boundary values.

Finally, we numerically studied the non-Markovian decoherence
control for three different conditions, i.e., $\omega_0\ll\omega_c$,
$\omega_0\approx\omega_c$, $\omega_0\gg\omega_c$ in the Ohmic
environment whose spectral density is with a Lorentz-Drude cuttoff
function. Our numerical results indicated that the decoherence
dynamics behaves differently for the different environmental
condition which leads to significant distinctness in the time
dependent behavior of the dissipation function $\gamma(t)$ and
$\Delta(t)$. We regarded temperature as a key factor in the
decoherence effect and showed that the decoherence can't be
controlled effectively in high temperature for both the Markovian
and non-Markovian. Comparing with the Markovian approximation we
believed that it is necessary to consider the non-Markovian quantum
system. Most of all, we analyzed the short time, moderate time, and
long time decoherence control behaviors for $r=0.1$, which implies
$\omega_c\ll\omega_0$. In this case the decoherence can be
controlled effectively, which may indicates that the decoherence
rate can be slowed down and decoherence time can be delayed through
designing some engineered reservoirs proposed by Myatt et. al.

\begin{acknowledgments}
This research is supported by the National Natural Science
Foundation of China (No. 60774099, No. 60221301) and by the Chinese
Academy of Sciences (KJCX3-SYW-S01). And the first author would like
to thank Dr. J. Zhang for many fruitful discussions.
\end{acknowledgments}

\subsection{\label{app:subsec}Comparing the non-Markovian dynamics with the Markovian dynamics }
\subsubsection{Quantum Markovian Process and Markovian Master Equation}
 Quantum Markovian process or Markovian approximation is
widely used in open quantum system, typically in interaction of
radiation with matter (weak coupling); quantum optics and cavity-QED
(weak damping); quantum decoherence; quantum Brownian motion (high
temperatures); quantum information; quantum error correction;
stochastic unravelling (Monte Carlo simulations); laser cooling
(L\'{e}vy statistics of quantum jumps), and so on. The essence of
quantum Markovian process contains three assumptions:

\begin{itemize}
\item{(i)}The initial factorization ansatz (Feynman-Vernon
approximation). At time $t=0$ the bath $B$ is in thermal equilibrium
and uncorrelated with the system $S$
\begin{equation}
 \begin{split}
\rho_{tot}(0)=\rho_{S}(0)\otimes\rho_{B};
 \end{split}
 \end{equation}
\item{(ii)}Weak system-bath interaction (Born approximation).
\item{(iii)}Markovian approximation. The relaxation time $\tau_{B}$ of
the heat bath is much shorter than the time scale $\tau_{R}$
($\tau_B\ll\tau_R$) over which the state of the system varies
appreciably.
\end{itemize}

Then it induced the dynamical map $\Phi_{t}$:
\begin{equation}
 \begin{split}
\rho_{S}(0)\rightarrow\rho_{S}(t)=\Phi_{t}\rho_{s}(0)=tr_{B}{[U_{t}(\rho_{S}(0)\otimes\rho_{B})U^{\dag}_{t}]}.
 \end{split}
 \end{equation}
With some conditions, like completely positive and Hermiticity and
trace  preservation we get a quantum dynamical semigroup:
$\Phi_{t}=exp[\mathcal {L}t],$ which implies the Markovian master
equation:
\begin{equation}
 \begin{split}
\frac{d}{dt}\rho_{s}(t)=\mathcal {L}\rho_{S}(t),
 \end{split}
 \end{equation}
 where generator of time evolution is in Lindblad form:
 \begin{equation}
 \begin{split}
\mathcal
{L}\rho_{S}(t)=-\frac{i}{\hbar}[H_{S},\rho_{S}]+\sum_{i}\gamma_{i}[a_{i}\rho_{S}a_{i}^{\dag}-\frac{1}{2}\{a^{\dag}_{i}a_{i},\rho_{S}\}].
 \end{split}
 \end{equation}

\subsubsection{Non-Markovian Dynamics and Non-Markovian Master Equation}
Non-Markovian dynamics system is not a new research problem, but
recently it received considerable consideration
\cite{H.P.Breuer,Maniscalco2,Breuer1,Breuer2,Breuer3}. Comparing
with the Markovian dynamics it has three properties: (i)Semigroup
property violated: slow decay of correlations, strong memory
effects; (ii)Initial correlations: classically correlated or
entangled initial states; (iii)Strong couplings and low
temperatures, with which we can studied the short-time behavior and
exact evolution of quantum decoherence. With the help of these three
properties we can derive effective equations (Master equations). As
far as we known, there are two ways to derive the master equation.
One is called the path-integral method by Halliwell et.al
\cite{Halliwell96},~Hu et. al \cite{Hu92},~Ford et. al\cite{ford01},
and Karrlein et.al\cite{karrlein}, the other is the projection
operator method by H.P.~Breuer \cite{H.P.Breuer, Breuer1, Breuer2,
Breuer3}.

The projection operator method is also called the Nakajima-Zwanzig
projection. The basic idea of the technique is to define a map
$\mathcal {P}$ as
\begin{equation}
\begin{split}
\mathcal {P}{\rho}=tr_{B}\{\rho\}\otimes\rho_B,
   \end{split}
\end{equation}
where $\rho_B$ is a fixed environment state and the map $\mathcal
{P}$ is a projection super-operator acting on operators, i.e.,
$\mathcal {P}^2=\mathcal {P}.$ Its complementary projection is
\begin{equation}
\begin{split}
\mathcal {Q}=\mathcal {I}-\mathcal {P},
   \end{split}
\end{equation}
where $\mathcal {I}$ is the identity map. Thus the Nakajima-Zwanzig
equation can be derived \cite{Breuer3}:
\begin{equation}
\begin{split}
\frac{d}{dt}\mathcal {P}\rho(t)=\int_0^tds K(t,s)\mathcal
{P}\rho(s)+\mathcal {I}(t)\mathcal {Q}\rho(0),
   \end{split}
\end{equation}
where $K(t,s)$ is the memory kernel. To second order in the coupling
constant the general form of the master equation can be approximated
by
\begin{equation}
\begin{split}
\frac{d}{dt}\mathcal {P}\rho(t)=\mathcal {K}(t)\mathcal
{P}\rho(t)+\mathcal {I}(t)\mathcal {Q}\rho(0).
   \end{split}
\end{equation}
In general, the TCL generator is

\begin{widetext}
\begin{equation}
\begin{split}
\mathcal
{K}(t)\rho_S=-\frac{i}{\hbar}[H_S(t),\rho_S]+\sum_i[C_i(t)\rho_SD_i^{\dag}(t)+D_i(t)\rho_SC_i^{\dag}(t)]-\frac{1}{2}\sum_i\{D_i^{\dag}(t)C_i(t)+C_i^{\dag}D_i(t),\rho_S\},
   \end{split}
\end{equation}
\end{widetext}
where $C_i(t)\neq D_i(t)$, which means that it is not in the
Lindblad form.

\end{document}